\documentclass[aps,pra,amsmath,amssymb,amsfonts,superscriptaddress,floatfix,nofootinbib]{revtex4} 
\pdfoutput=1
\usepackage{amsmath,amsthm,amsfonts,amssymb}
\usepackage{graphicx}
\usepackage{pythonhighlight}

\usepackage{color}
\usepackage{hyperref}

\usepackage{algorithm}
\usepackage[capitalize,nameinlink]{cleveref}

\newcommand{\be}{\begin{equation}}
\newcommand{\ee}{\end{equation}}

\newcommand{\cO}{{\cal O}}

\newcommand{\hc}{{\rm h.c.}}
\newcommand{\tp}{{\tilde \psi}}

\newcommand{\eps}{\epsilon}
\newcommand{\gs}{\Psi_0(\epsilon)}
\begin{document}

\title{Improving Perturbation Theory with the Sum-of-Squares: Third Order}
\author{Matthew B.~Hastings}
\begin{abstract}
The sum-of-squares method can give rigorous lower bounds on the energy of quantum Hamiltonians.  Unfortunately, typically using this method requires solving a semidefinite program, which can be computationally expensive.  Further, the typically used degree-$4$ sum-of-squares (also known as the 2RDM method\cite{coleman1963structure,erdahl1978representability,
percus1978role,mazziotti2001uncertainty,Nakata_2001,Maz12,klyachko2006quantum}) does \emph{not} correctly reproduce second order perturbation theory\cite{hastings2022perturbation}.  Here, we give a general method, an analogue of Wigner's $2n+1$ rule for perturbation theory, to compute the order of the error in a given sum-of-squares ansatz.  We also give a method for finding solutions of the dual semidefinite program, based on a perturbative ansatz combined with a self-consistent method.  As an illustration, we show that for a class of model Hamiltonians (with a gap in the quadratic term and quartic terms chosen as i.i.d. Gaussians), this self-consistent sum-of-squares method significantly improves over the 2RDM method in both speed and accuracy, and also improves over low order perturbation theory.  We then explain why the particular ansatz we implement is not suitable for use for quantum chemistry Hamiltonians (due to presence of certain large diagonal terms), but we suggest a modified ansatz that may be suitable, which will be the subject of future work.
\end{abstract}
\maketitle

\section{Introduction}
The sum-of-squares (SoS) method gives a way to prove \emph{lower} bounds on the energy of quantum Hamiltonians.  In quantum chemistry, this is known as the reduced density matrix method (RDM)\cite{coleman1963structure,erdahl1978representability,
percus1978role,mazziotti2001uncertainty,Nakata_2001,Maz12,klyachko2006quantum}.   These methods in fact form a hierarchy, with higher levels (or degrees) of the hierarchy being more accurate.

While this method can give powerful, nonperturbative results, unfortunately using the method in practice typically requires solving a semidefinite program (SDP).  It is very much \emph{not} clear whether general semidefinite programs can be solved in polynomial time without making some assumptions (see \cite{o2017sos} for some remarks on this in relation to the sum-of-squares).  Although  such assumptions likely hold for systems of practical interest, once we talk about practical systems we must also confront the issue that the running time of the SDP may be a relatively high-degree polynomial function of the system size, rendering it practically intractible for many systems.  Further, the degree of the polynomial will certainly increase as the degree of the sum-of-squares increases.

As a further problem, it was shown\cite{hastings2022perturbation} that the commonly used degree-4 sum-of-squares (or 2RDM) does not reproduce second-order perturbation theory for fermionic quantum Hamiltonians with quartic interactions, though degree-6 sum-of-squares does.
In this work, we consider this further.  We give a general rule, an analogue of Wigner's $2n+1$ rule, to understand errors in the sum-of-squares.  We then present an ansatz based on perturbation theory combined with a self-consistent method to not only show that degree-$6$ sum-of-squares can reproduce third order perturbation theory, but to rapidly \emph{find} a decomposition of the Hamiltonian as a sum-of-squares which certifies this result.  We then analyze this for a class of model Hamiltonians where it significantly outperforms the regular 2RDM in both speed and accuracy, and we discuss the further work needed to apply it to quantum chemistry.

This paper may be seen as a sequel to \cite{hastings2022optimizing,hastings2022perturbation}, and for that reason we omit giving some background which is standard in the field of the sum-of-squares, though some background, both to Wigner's rule and to the sum-of-squares itself, is in the next subsection.

\subsection{Wigner's Rule}
Wigner's $2k+1$ rule says that given a wavefunction which is accurate to order $k$ in perturbation theory, 
it can be used to compute the energy to order $2k+1$. 
This is usually called the $2n+1$ rule, but we will instead use the symbol $n$ to denote the number of fermionic modes in a many-body system and so from here on we call this the $2k+1$ rule.

A general setting for perturbation theory is to consider some Hamiltonian $H=H_0+\epsilon V$, where $H_0$ is some simple, ``unperturbed" Hamiltonian and where $\epsilon$ is a small parameter controlling the perturbation strength.
To understand Wigner's $2k+1$ rule, it is not necessary to have any deep understanding of a particular perturbation expansion, such as Rayleigh-Schr\"{o}dinger or Brillouin-Wigner.  Instead, it may be derived on general grounds.  Suppose $\psi^{(k)}$ is a normalized wavefunction with the property that
\be
\label{wigrule}
|H \psi^{(k)}-\lambda \psi^{(k)}| \leq \cO(\epsilon^{k+1}),
\ee
for some scalar $\lambda$,
such that $\lambda=\langle \psi^{(k)} | H |  \psi^{(k)} \rangle$, so that $\psi^{(k)}$ is an approximate eigenvector of $H$, with expectation value $\lambda$.
Suppose also that $\lambda$ is closer to some non-degenerate eigenvalue of $H$ than to any other eigenvalue, with that eigenvalue separated from the rest of the spectrum by an excitation gap (if $\epsilon$ is small, it suffices that $H_0$ has a unique ground state separated by an excitation gap and that $\lambda$ is close to that ground state energy).
We may imagine that $\psi^{(k)}$ is derived from some $k$-th order perturbation theory, but this is not necessary: the $2k+1$ rule will follow just from these assumptions.
Expand $\psi^{(k)}$ in an eigenbasis of $H$, with amplitudes $a_j$ for energy $E_j$.  If $\lambda$ is closest to some eigenvalue $E_0(\epsilon)$ of $H$, and all other eigenvalues $E_1(\epsilon),E_2(\epsilon),\ldots$ are separated by an excitation gap of order unity, then \cref{wigrule} implies that
$a_i=\cO(\epsilon^{k+1})$ for $i>0$.  Then, it follows that $\lambda= E_0+\cO(\epsilon^{2k+2})$.  That is, the error in energy is $\cO(\epsilon^{2k+2})$, so the energy is accurate to $\epsilon^{2k+1}$.

The sum-of-squares hierarchy for quantum systems is a method to prove lower bounds on the ground state energy of the Hamiltonian by
decomposing
\be
H=\sum_\alpha O_\alpha^\dagger O_\alpha + \lambda,
\ee
for some set of operators $O_\alpha$.
This proves a lower bound that the ground state energy, which we will write as $E_0(\epsilon)$, is greater than or equal to $\lambda$.
Indeed, taking $\gs$ to be the ground state wavefunction, we have
\be
E_0(\eps)-\lambda=\sum_\alpha |O_\alpha \gs|^2.
\ee
Hence, we immediately see that if $|O_\alpha \gs|=\cO(\epsilon^{k+1})$ for all $\alpha$, for some $k$, then $E_0(\eps)-\lambda=\cO(\epsilon^{2k+2})$, in analogy to the $2k+1$ rule for wavefunctions.  In words, if $\gs$ is approximately annihilated by each $O_\alpha$, up to error $\cO(\epsilon^{k+1})$, then sum-of-squares gives a lower bound on the energy which is tight to order $\epsilon^{2k+1}$, i.e., has error $\cO(\epsilon^{2k+2})$.
Conversely, if $|O_\alpha \gs| \gg \epsilon^{k+1}$ for some $\alpha$, then the sum-of-squares bound is \emph{not} accurate to the given order.

Our interest is in fermionic systems, with a Hamiltonian that is a polynomial in creation and annihilation operators.
The degree-$2k$ sum-of-squares hierarchy for this problem means that each $O_\alpha$ in the decomposition above may be up to degree-$k$ in the fermionic creation and annihilation operators.
The optimal $O_\alpha$ can be found by semidefinite programming.  The dual of this semidefinite program is a semidefinite program for \emph{pseudoexpectation} values of operators up to degree-$2k$.  Linear constraints are imposed on the pseudoexpectation values from the anticommutation relations.
Equivalently, one uses the anticommutation relations to show the equivalence of different ways of decomposing $H$ as a sum-of-squares; e.g., even though $H$ is only degree-$4$, one may write various different expressions which are polynomials of degree $\leq 2k$ which are equal to $H$.
These linear constraints can be computationally costly.  The degree-$2k$ sum-of-squares method requires then a semidefinite program with matrices of dimension $\Theta(n^k)$ and with $\Theta(n^{2k})$ constraints.

In general we will say that a particular degree of sum-of-squares reproduces $k$-th order perturbation theory if the bound on the energy from the optimum of the semidefinite program agree with the true ground state energy up to error $o(\epsilon^k)$.  Indeed, in all cases studied here, it will agree with the true ground state energy up to error $O(\epsilon^{k+1})$.

\subsection{Outline of Paper}
In \cref{third}, we show that degree-$6$ sum-of-squares reproduces third order perturbation theory for a Hamiltonian with quartic perturbation, and we give a self-consistent method to find a decomposition of the Hamiltonian.  This section serves also as a general explanation of the self-consistent method.
In \cref{numcon}, we show how to incorporate number and spin conservation into this method for increased speed of a numerical implementation.  In \cref{algo}, we slightly generalize the self-consistent method to improve numerical accuracy.  (A Python implementation of the algorithm can be found in \cref{implement}.)
In \cref{applications}, we test the algorithm on a class of model Hamiltonians, comparing to 2RDM and to perturbation theory, and we discuss the limitations of \emph{this particular implementation} of the self-consistent sum-of-squares method for quantum chemistry, where strong diagonal terms are present as well as moderate strength terms involving hopping a singlet pair of electrons from one orbital to another.
In \cref{extensions}, we discuss extensions of the self-consistent sum-of-squares method to handle this class of Hamiltonians in quantum chemistry and to be accurate to fifth and higher order in perturbation theory.  In \cref{dress}, we discuss some ``dressed" operators needed for these higher order results, and we consider size-consistency.

\section{Third Order Perturbation Theory}
\label{third}
We begin with reproducing third order perturbation theory by a sum-of-squares.
We consider a Hamiltonian
$H=H_0+\epsilon V$ with
$$H_0=\sum_i e_j \psi^\dagger_j \psi_j,$$
for some scalars $e_j>0$, where $\psi^\dagger_j,\psi_j$ are fermionic creation and annihilation operators, with a total of $n$ fermionic modes, and
where
$V$ is at most quartic in the fermionic creation and annihilation operators.
We assume further that $V$ is written as a normal ordered operator with vanishing scalar part (any scalar part can be incorporated into a shift in energy, of course).
Our assumption that all $e_j>0$ means that some particle-hole transformation may have been applied.  So, $V$ may not conserve fermion number even if the Hamiltonian before the particle-hole transformation did conserve fermion number.  In \cref{numcon} we discuss the conservation of fermion number and spin.

 In \cite{hastings2022perturbation}, it was proven that a certain ``fragment"\footnote{By a fragment of degree-$2k$ sum-of-squares, we mean including only some subset of terms possible terms in the sum-of-squares decomposition and including only some subset of the linear constraints in the semidefinite program, so that this is a semidefinite program which serves as an upper bound to the optimum of the degree-$2k$ sum-of-squares.} of degree-$6$ sum-of-squares reproduces second order perturbation but that degree-$4$ sum-of-squares does \emph{not} reproduce second order perturbation theory.  Further, it was conjectured that it degree-$6$ sum-of-squares reproduces third order perturbation theory.  We show in this section that this conjecture is correct.
 
Further, we give a self-consistent algorithm to find a sum-of-squares decomposition that does this.  Our basic idea is as follows: 
our goal is to find a sum-of-squares decomposition of some desired ``target Hamiltonian" $H_0+\epsilon V$.  To do this, we start with some
``low-order trial Hamiltonian" $H'$ that agrees with $H_0$ to low order in $\epsilon$, but may differ at higher order.
We use ideas from perturbation theory to give a guess at a sum-of-squares decomposition of that trial Hamiltonian.  
This sum-of-squares produces instead a sum-of-squares decomposition of some new Hamiltonian, that we call the ``effective Hamiltonian" $H_{eff}$, which may differ from the trial by some scalar plus some
higher order terms in $\epsilon$.  We start by setting the trial Hamiltonian equal to the target.  
If the difference between then effective Hamiltonian $H_{eff}$ and the target $H_0+\epsilon V$ is a scalar, we are done, as we have then found a sum-of-squares lower bound for the energy of the target.  However, if the difference is not a scalar, we shift the trial Hamiltonian by the non-scalar terms in the difference, and try again, repeating until the effective Hamiltonian equals the target Hamiltonian (or, in practice, is within some allowed tolerance of the target).

 We normal order the terms in $V$, meaning that the annihilation operators are to the right of creation operators.  Any quadratic term resulting after this normal ordering may be incorporated in $H_0$, so we assume indeed that $V$ is given to us as a normal ordered operator including only quartic terms.
 We
$V_{ijkl}$ to denote the coefficient of $\psi_i \psi_j \psi_k \psi_l$ in $V$.  
To denote terms with both creation and annihilation operators, we will
use overlines to denote creation operators, so that
$V_{\overline i \overline j k l}$ is the coefficient of $\psi^\dagger_i \psi^\dagger_j \psi_k \psi_l$, and similarly for terms with other numbers of creation and annihilation operators.
We will write coefficients always with the creation operators to the left of annihilation operators, and we will assume that $V_{abcd}$ is totally anti-symmetric in any set of indices which all have an overline or all do not have an overline.
So, we will write
\begin{align}
V=&\Bigl(\sum_{i<j<k<l} V_{ijkl} \psi_i \psi_j \psi_kj \psi_l+\hc\Bigr) + \Bigl(\sum_{i} \sum_{j<k<l} V_{\overline i j k l} \psi^\dagger_i \psi_j \psi_k \psi_l+\hc\Bigr)
\\ \nonumber &+ \sum_{i>j} \sum_{k<l} V_{\overline i \overline j k l} \psi^\dagger_i \psi^\dagger_j \psi_k \psi_l.
\end{align}
We emphasize that the above equation is indeed our choice of normalization of $V$; that is, there is no factor of $1/4!$ in the first term or $1/3!$ in the second or $(1/2!)^2$ in the third.

For use later, define $n_i=\psi^\dagger_i \psi_i$, and say that the number operator is equal to $\sum_i n_i$.  Say that a state has $k$ excitations if it is an eigenstate of the number operator with eigenvalue $k$.

From standard second order perturbation theory, denoting the unperturbed eigenvectors by $\Psi_j(0)$, with $\Psi_0(0)$ the unperturbed ground state, and with corresponding unperturbed energies $E_j(0)$,
we have
\begin{align}
\label{standard2pert}
\gs=&  \Psi_0(0)+\epsilon \sum_{j>0} \Psi_j(0) \frac{\langle \Psi_j(0) | V | \Psi_0(0) \rangle}{E_0(0)-E_j(0)}
+ \epsilon^2 \sum_{j,k>0} \Psi_j(0) \frac{\langle \Psi_j(0) | V | \Psi_k(0) \rangle}{E_0(0)-E_j(0)} \frac{\langle \Psi_k(0) | V | \Psi_0(0)\rangle}{E_0(0)-E_k(0)}
\\ \nonumber &-\epsilon^2  \sum_{j>0} \Psi_j(0) \frac{\langle \Psi_j(0) | V | \Psi_0(0) \rangle \langle \Psi_0(0) | V | \Psi_0(0)\rangle}{(E_0(0)-E_j(0))^2}
-\frac{1}{2}\epsilon^2 \Psi_0(0) \sum_{j>0} \frac{|\langle \Psi_j(0) | V | \Psi_0(0)\rangle|^2}{(E_0(0)-E_j(0))^2}+\cO(\epsilon^3),
\end{align}
where the last $\epsilon^2$ term is present as we normalize $|\gs|^2=1$.
We have assumed that the perturbation $V$ has vanishing scalar part when written in normal order, so
the second term of order $\epsilon^2$ vanishes as $\langle \Psi_0(0) | V | \Psi_0(0)\rangle=0$.
In fact, we will only need the perturbed wavefunction to order $\epsilon$, but we write down the wavefunction to order $\epsilon^2$ for use later.
Thus, to order $\epsilon$, $\gs=\Psi_0(0) + \sum_{i<j<k<l} \frac{1}{e_i+e_j+e_k+e_l} \psi^\dagger_i \psi^\dagger_j \psi^\dagger_k \psi^\dagger_l \Psi_0(0)+\cO(\epsilon^2)$.

To reproduce third order perturbation theory,
let us consider some \emph{other} Hamiltonian $H'$ given by
 $$H'=H'_0 + \epsilon V',$$
 where
 $$H'=\sum_i e'_i (\psi'_i))^\dagger \psi'_i,$$
 with the operators $\psi'$ being linear in the operators $\psi$ and $\psi^\dagger$ and with $\{\psi'_i,\psi'_j\}=0$ and $\{(\psi'_i)^\dagger,\psi'_j\}=\delta_{i,j}$.  Note that, for example, $\psi'_i$ may be a combination of $\psi$ and $\psi^\dagger$ say as $(1/\sqrt{2})(\psi'_i+\hc)$.
Assume $V'$ is written in the same way as $V$ in terms of some tensor $V'_{abcd}$.

This $H'$ is our low-order trial Hamiltonian.

We now define certain operators $\tau_i$ by
\begin{align}
\tau_i=\epsilon \sum_{j<k<l} T^{\overline i}_{\overline j\overline k \overline l} \psi^\dagger_j \psi^\dagger_k \psi^\dagger_l+
\epsilon \sum_{j<k} \sum_l T^{\overline i}_{\overline j\overline k l} \psi^\dagger_j \psi^\dagger_k \psi_l+
\epsilon \sum_{j} \sum_{k<l} T^{\overline i}_{\overline j k l} \psi^\dagger_j \psi_k \psi_l.
  \end{align}
Here the scalar $T$ is given by
\begin{align}
T^{\overline i}_{\overline j\overline k \overline l}=\frac{1}{e'_i+e'_j+e'_k+e'_l} V'_{\overline i \overline j \overline k \overline l}
+
\\ \nonumber
T^{\overline i}_{\overline j\overline k  l}=\frac{1}{e'_i+e'_j+e'_k} V'_{\overline i \overline j \overline k l}
+
\\ \nonumber
T^{\overline i}_{\overline j k  l}=\frac{1}{2}\frac{1}{e'_i+e'_j} V'_{\overline i \overline j k l}.
\end{align}

 Then,
 \be
 \label{issumofsq}
 \sum_i e'_i \Bigl( \psi'_i+\tau_i \Bigr)^\dagger \Bigl( \psi_i + \tau_i  \Bigr) +\sum_i e'_i \tau_i \tau_i^\dagger
=H' + \sum_i e_i \{ \tau_i,\tau_i^\dagger\}.
\ee

The left-hand side of \cref{issumofsq} is a sum-of-squares.  
The right-hand side is our effective Hamiltonian $H_{eff}$.
If the right-hand side is equal to $H-\lambda$ for some scalar $\lambda$, this gives a sum-of-squares proof that $H\geq \lambda$.

The term
$\sum_i e_i \{ \tau_i,\tau_i^\dagger\}$ is degree-$4$ in the Majorana operators.  
We normal order this term.  The scalar part after normal ordering arises only from considering terms with three creation operators in $\tau$ and terms with three annihilation operators in $\tau^\dagger$.
One may verify that 
\be
\sum_i e'_i \{ \tau_i,\tau_i^\dagger\} = W -\lambda,
\ee
for
\be
\lambda=-\epsilon^2 \sum_{i<j<k<l} \frac{|V'_{i,j,k,l}|^2}{e'_i+e'_j+e'_k+e'_l},
\ee
for some $W$, where $W$ is at most quartic (it may include quadratic terms, and these are the terms that may cause us to pick $H'_0$ different from $H_0$ below), and where $W$ is $O(\epsilon^2)$, and where $W$ has vanishing scalar part when written in normal ordered form.
(See \cite{hastings2022perturbation} for this calculation of the scalar term $\lambda$.  That reference considered the case that the quartic perturbation includes only terms with four creation operators or four annihilation operators but no mixed terms with both creation and annihilation operators.  However, the general calculation is similar.)
So, if we can pick $H'$ so that $H'+W=H$, then we have found a sum-of-squares proof that $H\geq \lambda$ as desired.

We first claim that, perturbatively in $\epsilon$, there is a choice of $H'$ which gives $H'+W=H$, where $W$ is implicitly a function of $H'$.  This may be shown iteratively.
The map $H'\rightarrow H'+W$ defines some function, that we call $f(H')$.  This function $f(H')$ is quadratic in $H'$.
Let us consider a sequence of choices of $H'$, denoted $H'(0),H'(1),H'(2),\ldots$.
Pick $H'(0)=H$.  
Let $$H'(j+1)=H'(j)+H-f(H'(j)).$$
Then, if $f(H'(j))-H$ is bounded by $\delta$ for some $\delta$ (meaning, for example, that it is a polynomial with all coefficients bounded by $\delta$), and if $V'$ is $O(\epsilon)$,
then
$$f(H'(j+1))-H=O(\epsilon \delta+\delta^2),$$
and hence tends to zero.

Of course, in practice, if one wishes to find a choice of $H'$ such that $f(H')=H$, other methods may be faster.
For example, effectively what we are doing is solving some nonlinear equation $f(H')=H_0+\epsilon V$ using an iterative method.  An iterative Newton method for solving this would involve computing the Jacobian of $f$, and using it to update $H'$ at each iteration, and so on.  Instead we are using a much simpler method to update $H'$ where we assume that the derivative of $f$ is given by its value at $H'=H_0$, i.e., we approximate the Jacobian by the identity matrix.  One could of course consider many other methods from nonlinear optimization such as conjugate gradient.

From this, one may see, without doing any calculation, that this self-consistent calculation correctly reproduces third order perturbation theory as follows.
 If we pick $\psi'_i=\psi_i$ and $V'=V$ then, from the perturbative wavefunction \cref{standard2pert} above,
 we have $(\psi_i+\tau_i)\Psi_0(\epsilon)=O(\epsilon^2)$.
 We have $\psi'_i=\psi_i+O(\epsilon^2)$ and similarly $V'=V+O(\epsilon)$ from the iterative solution\footnote{This is an $O(\epsilon)$ difference between $V'$ and $V$, so an $O(\epsilon^2)$ difference in the corresponding Hamiltonians.}, so also for whatever the given final $V'$ is we have
 $(\psi'_i+\tau_i)\Psi_0(\epsilon)=O(\epsilon^2)$.
 Similarly, one may verify that $\tau_i^\dagger \Psi_0(\epsilon)$ is $O(\epsilon^2)$.
 So, by the $2k+1$ rule, indeed the error in energy is $O(\epsilon^4)$.
 
 However, it is useful to do an explicit calculation to verify third order perturbation theory.
 We have $H'(0)=H$.  We find that $H'(1)-H=\cO(\epsilon^2)$ .  
 We will write $V'(j)$ for the tensor determining the quartic terms in $H'(j)$.
 If we consider just the terms with four annihilation operators in this difference, then we find that
 $V'(1)_{ijkl}-V_{ijkl}$ is determined by terms in $W$ with four annihilation operators, which in turn arise from the sum over $m$ of $e_m$ times the anti-commutator of a term in $\tau_m^\dagger$ with a term in $\tau$, where the term in $\tau_m^\dagger$
 has
 three annihilation operators and the term in $\tau_m$ has two annihilation operators and one creation operator.
 These terms in $\tau_m$ and $\tau_m^\dagger$ in turn arise from a term in $V$ with
four annihilation operators and a term in $V$ with two annihilation operators and two creation operators, respectively.
We find that
 $$V'(1)_{ijkl}= V_{ijkl}-
 \epsilon \frac{1}{2}\sum_{m,n} e_m  \frac{V_{ijmn}}{e_i + e_j + e_m + e_n} \frac{V_{\overline m\overline nkl}}{e_m+e_n}.$$
 This in turn leads to a change in the second order perturbation theory energy, so that the second order perturbation energy of $H'(1)$
 is equal to that for $H'(0)=H$ plus
 $$\epsilon^3\frac{1}{2}\Bigl(\sum_{i<j<k<l} e_m \sum_{m,n} \frac{V_{\overline i\overline j\overline k\overline l}}{e_i+e_j+e_k+e_l}
 \frac{V_{ijmn}}{e_i + e_j + e_m + e_n} \frac{V_{\overline m\overline nkl}}{e_m+e_n}
 \Bigr) +\cO(\epsilon^4).$$
 Note the sign: the second order perturbation correct has a negative sign in front, but then there is an additional negative sign in front since we have \emph{subtracted} something from $V$ to find $V'(1)$, so the overall sign is positive.
 Note also that there is no longer a factor of $1/2$: the change in $V'(1)$ had a factor of $1/2$, but the expression for the second order perturbation theory is quadratic in $V$, so we get a factor of $2$ from differentiating a quadratic function.
 
 The sum over $m,n$ is unrestricted, but of course the term vanishes if $m=n$.  We can restrict the sum to $m<n$ and add in a summand with $m,n$ interchanged.  In this way, we replace $e_m$ with $e_m+e_n$, which then cancels against $e_m+e_n$ in the denominator.
 This simplifies the expression to
  $$\epsilon^3\Bigl(\sum_{i<j<k<l} \sum_{m<n} \frac{V_{\overline i\overline j\overline k\overline l}}{e_i+e_j+e_k+e_l}
 V_{\overline m\overline nkl} \Bigr) \frac{V_{ijmn}}{e_i + e_j + e_m + e_n} 
 +\cO(\epsilon^4),$$
 which may be recognized as the standard third order perturbation theory correction to the energy.
 If we consider $H'(2),H'(3),\ldots$, this does not change $V'$ to order $\epsilon$ and so gives the same third order perturbation energy.

Now consider a specific example to see how this works at strong coupling.
Let
\be
\label{mod1H}
H=\sum_{i=1}^4\psi^\dagger_i \psi_i+\epsilon \Bigl( \psi^\dagger_1 \psi^\dagger_2\psi^\dagger_3 \psi^\dagger_4+\hc\Bigr).
\ee
The ground state of this model is a linear combination of state $|0\rangle$ and $|4\rangle$, where $|0\rangle$ has no particles present and $|4\rangle$ has four particles present so that $\psi_i|0\rangle=0$ and $\psi^\dagger_i|4\rangle=0$.
Strong coupling is the case $\epsilon \gg 1$.
We pick $\psi'_i=\psi_i$ and $e_i=e$ for some $e$, choosing
$\tau_i=a \prod_{j\neq i} \psi^\dagger_i + b \psi_i \sum_{j \neq i} n_j,$
where $n_j=\psi^\dagger_j \psi_j$ and
where the ordering of terms in the product is chosen so that the coefficient proportional to $a$ in $\psi^\dagger_i \tau_i$ is $\psi^\dagger_1 \psi^\dagger_2 \psi^\dagger_3\psi^\dagger_4$.
Such a choice of $\tau_i$ can arise from repeating the self-consistent procedure above.

Then, the left-hand side of \cref{issumofsq} is equal to
$$\Bigl(3 eb^2 - 3 e a^2 + e\Bigr) \sum_i n_i
+\Bigl(4ea+12eab\Bigr) \Bigl( \psi^\dagger_1 \psi^\dagger_2+\psi^\dagger_3 +\psi^\dagger_4+\hc\Bigr)
+\Bigl(2eb+ea^2+2eb^2\Bigr) \sum_{i \neq j} n_i n_j
-\lambda,$$
for $\lambda=-4ea^2$.
We must have $2eb+ea^2+2eb^2=0$ and $4ea+12eab=\epsilon$ and $3 eb^2 - 3 e a^2 + e=1$.
After some algebra, one finds that in this case, we have, for any $\epsilon$, that
$\lambda=-\epsilon^2/4$.  That is, for \emph{any} $\epsilon$, we obtain exactly the second order perturbation theory result, so this clearly does not work well for $\epsilon\gg 1$.   We emphasize that the equations have real solutions for all real $\epsilon$.

We can see why this method does not correctly give the behavior at $\epsilon \gg 1$, because the term $\sum_i e'_i \tau_i \tau_i^\dagger$ is not small.
This then suggests how to generalize the sum-of-squares that we consider to get the correct strong coupling.
We simply consider a decomposition, for general H,
\begin{align}
\label{moregen3}
H&=
 \sum_i \Bigl( \theta_i+\tau_i \Bigr)^\dagger \Bigl( \theta + \tau_i  \Bigr) +\sum_i \Bigl(\xi_i+ \tau\Bigr) \Bigl(\xi_i+\tau_i\Bigr)^\dagger+\lambda\\ 
 \nonumber
 &=\sum_i \theta_i^\dagger \theta_i + \Bigl(\xi_i \xi_i^\dagger + \theta_i^\dagger \tau_i + \hc \Bigr)+ \Bigl( \xi_i \tau_i^\dagger +\hc \Bigr) + \{\tau_i,\tau_i^\dagger\}+\lambda
 ,
 \end{align}
  for some choice of $\tau_i$,
 where $\theta_i,\xi_i$ are linear in the fermion operators (in the previous case, we took $\xi_i=0$ and $\theta_i=\sqrt{e_i} \psi'_i$).
 For now we are \emph{not} considering applying the self-consistent method to this, but simply asking how well this decomposition can bound the ground state energy for optimal choice of $\tau_i,\theta_i,\xi_i$.

For this particular
example, we pick
\begin{align}
\label{mod1}
H=&\sum_i \Bigl(\psi_i + a \prod_{j \neq i} \psi^\dagger_j + b \psi_i \sum_{j \neq i} n_j\Bigr)^\dagger \Bigl(\psi_i + a \prod_{j \neq i} \psi^\dagger_j + b \psi_i \sum_{j \neq i} n_j \Bigr)
\\ \nonumber
&
+s\sum_i \Bigl( b \psi_i^\dagger \sum_{j\neq i} n_j + a \prod_{j \neq i} \psi_i+\frac{a^2}{1+3b} \psi_i^\dagger\Bigr)|^\dagger
\Bigl( b \psi_i^\dagger \sum_{j\neq i} n_j + a \prod_{j \neq i} \psi_i+\frac{a^2}{1+3b} \psi_i^\dagger \Bigr),
\end{align}
where the ordering of terms in the product is as before.  The coefficients of terms are chosen so that the first and second sums in \cref{mod1} both vanish on the same linear combination of $|0\rangle$ and $|4\rangle$, so that this decomposition will give the exact ground state energy of the resulting Hamiltonian and that linear combination must be the ground state.
We have found numerically that it is possible to choose
$a,b,s$ so that the Hamiltonian of \cref{mod1} is equal to $\mu \sum_i n_i+ J(\psi_1 \psi_2 \psi_3 \psi_4+\hc)$, up to a scalar, so long as 
$|J/\mu|$ is smaller than $\approx 9.7$, i.e., it works up to rather large coupling.

Of course,
there are other ways to decompose such a simple Hamiltonian as a sum-of-squares.  The nontrivial thing is that we have done this using terms of odd fermion parity, with a particular relation between the odd fermion parity terms of degree-$3$ present, so that terms of degree-$6$ in the sum-of-squares appear only as anticommutators.

Of course, our choice of $\psi'_i,e'_i,\tau_i$ such that  the sum-of-squares is equal to $H$ plus a scalar as non-unique.  For example, just consider the quadratic terms in $H$.  If two energies agree, e.g. $e_1=e_2$, then one may take $\psi'_1,\psi'_2$ to be any orthogonal combination of $\psi_1,\psi_2$.  Even if the energies do not agree, if we had not imposed that $\{\psi'_i,\psi'_j\}=0$ and $\{(\psi'_i)^\dagger,\psi'_j\}=\delta_{i,j}$, the choice would be non-unique.  We can see this by counting parameters.  A quadratic $H_0$ of the form $\sum_{ij}\psi^\dagger_i H_{ij} \psi_j$ is determined by $n(n+1)/2$ real parameters if there are $n$ fermionic modes, while
there are $n^2$ real parameters determining possible choices of $\psi'_i$ which are linear combinations of $\psi_i$.
One may see a similar non-uniqueness by counting parameters in possible choices of $\tau$ compared to the parameters in $V$.
However, the particular form we have chosen for $\tau$ seems to be a useful one, and also, by restricting to this form, it simplifies any numerical application of the procedure.  If we allowed more general choices of $\tau$, then, while it may give better bounds on energy, it would also be more numerically costly.

\section{Number and Spin Conservation}
\label{numcon}
For efficient implementation of these methods, it is useful to be able to include spin conservation and particle number conservation.
We write fermionic anihilation operators as $\psi_{i,\sigma}$ where $i$ indexes different orbitals, with $1\leq i \leq n_o$, with $n_o$ being the total number of orbitals, and where $\sigma\in \{\uparrow,\downarrow\}$ denotes the spin degree of freedom.  Let $n_i=\sum_{\sigma} \psi^\dagger_{i,\sigma} \psi_{i,\sigma}$ be the number operator.

We assume that all terms in the Hamiltonian obey particle number conservation, so that they have an equal number of creation and annihilation operators.  Thus, we can write the quadratic term in the Hamiltonian as
$$H_0=\sum_{i,j} \sum_{\sigma} (h_0)_{i,j} \psi^\dagger_{i,\sigma} \psi_{j,\sigma},$$
for some matrix $h_0$.
Since we have \emph{not} done a particle-hole conjugation (as was done earlier so that all energies $e_i$ in $H_0$ would be positive),
the matrix $h_0$ may have both positive and negative eigenvalues.

We will write the quartic term in the Hamiltonian as
\be
\label{Vconserve}
V=\sum_{i,j,k,l} \sum_{\sigma,\tau} G_{ijkl} \psi^\dagger_{i,\sigma} \psi_{j,\sigma} \psi^\dagger_{k,\tau} \psi_{l,\tau},
\ee
for some four-index tensor $G$.

One may wonder: here we have constructed a term, $V$, which has spin $0$, as a sum of products of two terms (i.e.,  
$ \psi^\dagger_{i,\sigma} \psi_{j,\sigma}$ and $ \psi^\dagger_{k,\tau} \psi_{l,\tau}$) which themselves are spin $0$, but
are there other ways to construct such a spin-$0$ $V$?  The reader may indeed verify that any choice of $V$ with total spin $0$ can be written as a sum of a term of this form plus possibly some quadratic term.

Note that the Hermitian conjugate of $V$, i.e, $V^\dagger$, may be obtained by replacing $G_{ijkl}$ with $G_{lkji}$.  Note also that even if $G_{ijkl}$ is different from $G_{lkji}$, then the Hamiltonian $H_0+V$ may still be Hermitian!  This is because if we replace $G_{ijkl}$ with $G_{klij}$ then this changes $V$ by adding a quadratic term; thus, it is possible that we have a non-Hermitian $V$ and a non-Hermitian $H_0$, but the sum is still Hermitian.
In the algorithm later, we will explicitly keep $H_0$ and $V$ both Hermitian so that $G_{ijkl}=G_{lkji}$.

To exploit number and spin conservation, the operators $\tau$ will also have a spin index.  We will make operators $\tau_{i,\sigma}$ always have charge $-1$(i.e., they destroy a single particle, the same as $\psi_i$), and will always be spin-$1/2$.   Such an operator, of degree at most $3$, can then always be written as a sum
$$\tau_{i,\sigma}=\sum_{j,k,l} \sum_\tau T^i_{j,k,l} \psi_{j,\sigma} \psi^\dagger_{k,\tau} \psi_{l,\tau}+\textrm{linear term},$$
for some four-index tensor $T$, where the linear term is some linear combination of annihilation operators with spin index $\sigma$.
This four-index tensor is stored in the correspondingly named variable $T$ in the Python code later, while the linear term is stored in the matrix $T1$.

With this definition of $\tau_{i,\sigma}$, if eigenvalue $e_i$ is positive, then this is the same definition of $\tau_{i,\sigma}$ as before, but if eigenvalue $e_i$ is negative, this is the hermitian conjugate of the previously defined $\tau_{i,\sigma}$.

So, since we have not done particle-hole conjugation, some terms in the sum-of-squares will be of the form
$(\psi^\dagger_i +\epsilon \tau^\dagger_i)(\psi_i + \epsilon \tau_i)$ if the $i$-th eigenvalue is positive, and others will be of the form
$(\psi +\epsilon \tau)(\psi^\dagger_i + \epsilon \tau^\dagger_i)$ if the $i$-th eigenvalue is negative.

Given this particular choice of $\tau$, after some algebra one may express an anti-commutator $\{\tau_i^\dagger,\tau_j\}$ as a combination of scalar, quadratic, and quartic terms, all obeying spin and number conservation, so that the quadratic and quartic terms are of the same form as $H_0$ and $V$ above, respectively.  The result of this algebra is included in some of the einsum commands in the Python program later.

\section{Generalized Algorithm}
\label{algo}
The decomposition of the Hamiltonian above encounters an issue.  If we write $V$ as in \cref{Vconserve}, then $V$ is \emph{not necessarily} normal ordered.  As a result, if one normal orders $V$, then additional quadratic terms might appear.  Indeed, the appearance of these terms might indicate that the ground state of $H_0$ is not the Hartree-Fock ground state.

To resolve this, we use an additional ``guide Hamiltonian".  This guide Hamiltonian is a quadratic Hamiltonian.  When we compute the effective Hamiltonian from the low-order trial Hamiltonian, previously we assumed that the quadratic piece of the trial Hamiltonian determined a set of eigenvectors and corresponding eigenvalues $e_i$.  Those eigenvalues were used to compute the $\tau_i$.  Now, we use the eigenvectors and eigenvalues of the guide Hamiltonian, rather than the quadratic part of the trial Hamiltonian, to determine the $\tau_i$.

We update the guide Hamiltonian iteratively through the run as explained below.

A second issue that we encounter is purely an issue of convenience: all of the operators $\tau_i$ that we define annihilate a single particle, but some terms in the sum-of-squares involve positive eigenvalue and some involve negative, as explained above.  To simplify figuring out the decomposition as a sum-of-squares, we do the following: first, given the guide Hamiltonian and trial Hamiltonian, we construct operators $\tau_i$ which involve both linear and cubic terms as explained above.  
We then use the $\tau_i$ to write the trial Hamiltonian as a bilinear in the operators $\psi^\dagger, \psi,\tau^\dagger,\tau$.  That is, we exactly decompose the quartic terms in terms of $\psi^\dagger \tau$ terms (and their Hermitian conjugate) as well as possibly $\psi^\dagger \psi$ terms.
We call this bilinear $H_{\psi \tau}$.

We then find a sum-of-squares decomposition of $H_{\psi\tau}$, up to $\tau^\dagger \tau$ terms.  To find this decomposition, we proceed by using an approximation that the the anticommutator of operators $\tau^\dagger,\tau$ with each other or with $\psi$ or $\psi^\dagger$ is some scalar, rather than some more complicated operator.  This scalar we define by computing the expectation value of the anticommutator in the ground state of the guide Hamiltonian.
This scalar is encoded in some $2n_o$-by-$2n_o$ ``anti-commutator matrix", describing the anticommutator of either a $\psi^\dagger$ or $\tau^\dagger$ with a $\psi$ or $\tau$.  This anti-commutator matrix is stored in the variable $Stemp$ in the Python code later.
If this approximation were exact, to find the optimum sum-of-squares decomposition, one first finds a new basis of $\psi$ and $\tau$ operators which diagonalizes this anti-commutator matrix.  Indeed, we choose this transformation to be non-orthogonal so that the anti-commutator matrix becomes the identity matrix in the new basis.  Then, one can rewrite $H_{\psi\tau}$ in this new basis, diagonalize the resulting matrix, and write the sum-of-squares where each term in the sum-of-squares corresponds to some eigenvector.
Remark: this could also be written as a general eigenvalue problem involving $H_{\psi\tau}$ and $Stemp$.

The error in this approximation (that the anticommutator in fact may include quadratic and quartic terms) leads then to the effective Hamiltonian differing from the trial Hamiltonian by quadratic and quartic terms, which then gets iterated as explained before.

In this procedure, we compute a new guide Hamiltonian by taking the $n_o$-by-$n_o$ subblock of $H_{\psi\tau}$ which includes only $\psi^\dagger \psi$ terms.

Let's discuss the complexity of this method.  In our previous algorithm of \cref{third}, the runtime of each iteration is dominated by finding the anti-commutators $\{\tau^\dagger,\tau\}$.  There were $n_o$ different anticommutators to compute, and each anticommutator can be computed in time $O(n_o^5)$ as each anti-commutator includes quartic terms with $O(n_o^4)$ elements, each of which can be computed in time $O(n_o)$.  So, the time per iteration is $O(n_o^6)$, the same complexity as to compute third order perturbation theory.
Of course, we can more directly express the computation of these anti-commutators as a certain matrix multiplication of $O(n_o^2)$-by-$O(n_o)$ matrices, giving the same time complexity.  Remark: here we ignore the possibility to speed this up by using fast matrix multiplication techniques; such a speedup could of course be applied to the calculation of third order perturbation theory too.

Naively, the method here takes time $O(n_o^7)$ per iteration as the there are $O(n_o^2)$ anti-commutators to compute.  That is also the time taken per iteration by the Python code later.  However, it may be possible to speed this to $O(n_o^6)$ per iteration as follows.  If we knew the basis which diagonalize the anti-commutator matrix, we could directly compute the anti-commutators in this basis, taking only time $O(n_o^6)$ per iteration.  One may be able to obtain an approximation to this basis on some small number of iterations (or perhaps recompute it every few iterations).  We leave this for the future.

In comparison, there in fact are few theoretical bounds for the performance of a semidefinite programming solver.  Indeed, without knowing some additional properties, one cannot claim that a general semidefinite program can be solved in polynomial time.  In practice, though, we have found that this iterative method is faster for the implementations we tried.

A final difference in the method of the Python implementation later, compared to that in \cref{third}, is that having computed a new effective Hamiltonian (and guide Hamiltonian), rather than directly replacing the trial Hamiltonian by the trial Hamiltonian shifted by the difference between the effective and target Hamiltonian, we instead shift only by some scalar constant times that difference.  This is the quantity $f$ in the Python program.  This seems to improve the convergence radius at large interaction, but slows the convergence at small interaction, and a better implementation may change this from iteration to iteration.

\section{Applications and Tests}
\label{applications}
Here we now give some brief tests of this method, showing that in some cases it is both faster and more accurate than certain implementations of the degree-$4$ sum-of-squares method, also called the RDM.  The results here are \emph{not} intended to be a detailed comparison.  Both methods (this method and the 2RDM) were likely implemented in a non-optimal fashion, and one may, for example, change the convex solver used in the 2RDM or also change the method of solving the self-consistent equations here.  Rather than a precise comparison, this is intended purely to give motivation to consider this method further.

\subsection{Some Model Hamiltonians}
To test this self-consistent method, we took $n_0=6$ and took
\be
H_0=\sum_{i=1}^{n_o/2} n_i - \sum_{i=n_o/2+1}^{n_0} n_i,
\ee
so that there were an equal number of positive and negative energies\footnote{The only reason we took such a small $n_o$ is that we used a very simple, nonsparse solver for the exact energy; the implementation of the method here and the 2RDM both run much faster than this solver even at this size.}

We choose the entries of $G_{ijkl}$ by $G_{ijkl}=(1/\sqrt{2})(A_{ijkl}+A_{lkji})$ where each $A_{ijkl}$ is an independent normal random variable with mean zero and variance $1$.  In this way, $V$ is Hermitian.

We have $H=H_0+\epsilon V$.
Numerically, we have found that the most negative eigenvalue of $V$ is, on average, approximately equal to $70$ in absolute value.  The most negative eigenvalue of $H_0$ is, of course, equal to $-n_0=-6$.  So, the eigenvalues become the same in average magnitude for $\epsilon$ slightly smaller than $0.1$.
We have tested the self-consistent procedure described above, and in \cref{implement}, for $\epsilon=0.01,0.02,0.04$.

We found the exact ground state energy, and computed the difference between the lower bound on energy found by the above method and the exact energy.  We also compared it to the difference from the 2RDM method.  For the 2RDM method\footnote{Our implementation is based on work with D. Wecker.}, we included spin and number conservation, and also included a constraint on total number (which breaks size-consistency as described in \cref{sizecon} and hence gives some advantage to the 2RDM method in reducing error for these small sizes).

In terms of error, namely the difference between the exact ground state energy and the bound on ground state energy produced either by the self-consistent method or 2RDM, the self-consistent method was consistently better.  At $\epsilon=0.01$, the average, of the ratio of the error of the self-consistent method to the error of the 2RDM, was equal to $\approx 0.012\ldots$ over $100$ runs.  The self-consistent method was also consistently faster, with the average, of the ratio of time taken by the self-consistent method to that taken by the 2RDM for a given instance,  being equal to $\approx 0.11$.
At $\epsilon=0.02$, the average of the ratio of errors was $\approx 0.048\ldots$, with the average time ratio being $\approx 0.12$.

   At $\epsilon=0.04$, we began to encounter convergence issues.  
   One way to diagnose these convergence issues is how the difference between the desired Hamiltonian and the sum-of-squares found by the iterative procedure.
   Measuring this error as a sum of matrix elements,
      for only $46$ out of $100$ runs was it able to converge so that this error was less than $10^{-4}$.  For these runs, however, the average ratio of errors was $\approx 0.12$ and the average time ratio was $\approx 0.15$.
We expect that the convergence of the self-consistent method could be improved by using something other than this simpler iterative procedure.  Perhaps once when one is close, one can use conjugate gradient methods or other methods of solving linear equations to speed convergence, or perhaps other numerical algorithms can be used for solution to better deal with nonlinear effects.

We also tested some runs with $n_o=20$.  In this case, we do not have exact results to compare to.  However, runs at $\epsilon=0.001,0.002,0.004$ showed better bounds with the self-consistent method than the 2RDM (i.e., even though we do not know the exact ground state energy, we know that the bound is better as the bound is more positive), and showed speedups ranging from $40$ times faster to $60$ times faster, with, surprisingly, some larger speeds occuring at larger $\epsilon$.  Thus, the speedup seems to increase at larger sizes.

Comparisons to first, second, and third order Rayleigh-Schrodinger perturbation theory also showed that the self-consistent method was almost always more accurate, often significantly more accurate, even at fairly weak couplings.  We omit any precise measurement of how much more accurate, but, for example, at $n_o=4$ and $\epsilon=0.01$, the average of the natural log of the error of the self-consistent method divided by the log of the error in third order Rayleigh-Schrodinger was observed numerically to be roughly $-1.9\ldots$.
  Of course, also the self-consistent method provides a strict lower bound, while the result from perturbation theory is not necessarily an upper or lower bound, so we define the error for Rayleigh-Schrodinger as the absolute value of the difference to the ground state energy.

Of course, other semidefinite programming solvers may be faster.  At the same time, likely the self-consistent method could be sped up.  At minimum, it is likely that one could adaptively adjust the quantity $f$ in the Python code which determines how large a step is taken at each iteration, but likely other numerical methods would speed up the solution.  We are using CVXPY which uses SCS as a semidefinite programming solver, and using numpy's einsum to compute tensor contractions for the self-consistent method, and perhaps both of these may be improved.
Real Hamiltonians may also allow speedups due to structure as, for example, $V$ may have some low-rank decomposition, but likely these speedups could be incorporated here.

\subsection{Quantum Chemistry}
Unfortunately, when applied to some small example molecules from quantum chemistry, this implementation of the self-consistent method does not perform well, encountering either convergence issues, poor accuracy, or both.
We discuss the reasons for this here.  In \cref{extensions}, we discuss plans to deal with these issues.

The issue is that the tensor $G$ describing the interaction is very different from the model Hamiltonians above.  Rather, the largest coefficients correspond to density-density interactions, i.e., terms of the form
$$n_i n_j = \sum_{\sigma,\tau} \psi^\dagger_{i,\sigma} \psi_{i,\sigma} \psi^\dagger_{j,\tau} \psi_{j,\tau}.$$
The next largest (smaller than those above, but still larger than others) correspond to
hopping of a singlet from orbital $j$ to $i$, i.e.,
$$\sum_{\sigma,\tau} \psi^\dagger_{i,\sigma} \psi_{j,\sigma} \psi^\dagger_{i,\tau} \psi_{j,\tau},$$
or to spin-spin interaction, i.e.,
$$\sum_{\sigma,\tau} \psi^\dagger_{i,\sigma} \psi_{j,\sigma} \psi^\dagger_{j,\tau} \psi_{i,\tau},$$
or to density-controlled hopping, i.e.,
$$\sum_{\sigma} \psi^\dagger_{i,\sigma} \psi_{j,\sigma} n_k.$$

To understand the issue with the density-density interactions, 
note that we have shown above that the self-consistent method reproduces third order perturbation theory, perturbing \emph{in the quartic terms}, but if we keep the density-density interactions at some fixed nonzero strength then this method does not reproduce third order perturbation theory in the remaining quartic terms.  To see this, let us absorb the factor of $\epsilon$ into the remaining terms (i.e., those other than the density-density interaction), so that $H=H_0+V$ rather than $H=H_0+\epsilon V$, where $V$ is still quartic and $H_0$ is still quadratic.  Then, we get some decomposition of the Hamiltonian as a sum-of-squares, with terms of the form $(\psi_i^\dagger+\tau_i^\dagger)(\psi_i+\tau)$ and of the form $\tau_i \tau_i^\dagger$, so we have also absorbed the factor of $\epsilon$ into $\tau_i$.  This operator $\tau_i \tau_i^\dagger$ is now an operator of order unity rather than of order $\epsilon^2$, because the density-density interaction is of order unity.  Then, its expectation value in the ground state is order $\epsilon^2$ rather than, as we had previously, order $\epsilon^4$; the reason it is order $\epsilon^2$ now is that the expectation value vanishes in the unperturbed ground state as the number operator is equal to zero there.  A separate issue is that the operator $\psi_i+\tau_i$ will not annihilate the ground state up to error $O(\epsilon^2)$, because our definition of $\tau_i$ assumed that the energy of the excited states was given (up to error $O(\epsilon)$) by the sum of single particle energies but now if density-density interactions are strong then this is not true; this separate issue is more easily solvable because we can choose to define $\tau_i$ by incorporating the density-density interaction into the energy of the unperturbed states so that $\psi_i+\tau_i$ is still at most cubic and so that it annihilates the ground stdate up to error $O(\epsilon^2)$.

Later we discuss some approaches to deal with the density-density terms.  As for the singlet hopping terms, it is worth noting that the 2RDM can solve the simplest case of this exactly, i.e., when we have two orbitals, one with positive energy and one with negative, and we have singlet hopping between them.  
That is, $H=n_0-n_1+\epsilon\sum_{\sigma,\tau}  (\psi^\dagger_{0,\sigma} \psi_{1,\sigma} \psi^\dagger_{0,\tau} \psi_{1,\tau}+\hc).$
In fact, 2RDM can solve this exactly even if we add a term $Un_0n_1$ to the Hamiltonian.\footnote{However, it is interesting to understand \emph{how} 2RDM solves this exactly in this case even if we do not impose fixed particle number.  If $U>0$, then we can use that 2RDM proves that $Un_0n_1>0$ so that 2RDM can lower bound by the energy in the case that $U=0$, and since the ground state has $n_0n_1=0$ then this lower bound is correct.  However, if $U<0$, then instead we can use a decomposition as a sum-of-squares using terms $O^\dagger O$ for $O=\psi^\dagger_{0,\sigma} \psi_{1,\sigma}+a\psi^\dagger_{1,\sigma}\psi_{0,\sigma}$ for an appropriate choice of $a$.  It is also worth noting that 2RDM does \emph{not} give the correct answer if we do not impose fixed particle number for a related Hamiltonian with $n_o=3$; this Hamiltonian is $H=n_0-n_1-n_2+\epsilon\sum_{\sigma,\tau}  (\psi^\dagger_{0,\sigma} \psi_{1,\sigma} \psi^\dagger_{0,\tau} \psi_{1,\tau}+\hc+1\leftrightarrow 2)$.}
However, the self-consistent method, as implemented below, does not exactly solve this model, though it gets a fairly accurate answer.  The reader may be surprised that the method does not exactly solve this model, as in \cref{third} we showed that the Hamiltonian of \cref{mod1H}, which is essentially the same problem, can be solved exactly for some range of coupling.  The issue is that our self-consistent solution does not find the optimum decomposition given in \cref{mod1}.  We have tried artificially modifying the variable $Stemp$ in the Python code below in some numerical experiments, so that it chooses a different decomposition, and have found in many cases that this significantly improves the accuracy.  It would be interesting to investigate whether there is some general method to improve the accuracy in this way for the kind of decomposition we consider here, and whether such a method would require solving a semidefinite program or not (possibly, it would require only a smaller program using matrices of size $O(n_o)$-by-$O(n_o)$) as we only have $O(n_o)$ variables $\psi,\tau$ that we consider).

We have also considered some model Hamiltonians with larger $n_o$, with all possible singlet hopping terms present, choosing the coefficients of these terms as independent, identically distributed Gaussians.  In some parameter regimes, we observe improvements in both speed and accuracy using the self-consistent method compared to the 2RDM.

\section{Discussion and Extensions}
\label{extensions}
It would be interesting to extend this method to reproduce fifth and higher order perturbation theory.  One approach is the following.  We have considered Hamiltonians $H=H_0+\epsilon V$, where $V$ is quartic, and our low order trial Hamiltonian was constructed such that it gave a sum-of-squares which is equal to $H$, plus some scalar term, up to error $O(\epsilon^2)$ and so that each term in the sum-of-squares had expectation value $O(\epsilon^4)$ in the ground state.  Thus, to reproduce fifth order perturbation theory, a natural choice is to consider a family of Hamiltonians $H=H_0+\epsilon V^{(4)} + \epsilon^2 V^{(6)}$, where $V^{(4)}$ is quartic and $V^{(6)}$ is sixth order, i.e., even if we are only interested in finding sum-of-squares decompositions of Hamiltonians with $V^{(6)}=0$, one should consider trial Hamiltonians from this more general family with possibly nonzero $V^{(6)}$.
Given such a trial Hamiltonian, we would then construct a sum-of-squares using perturbation theory so that each term in the sum-of-squares has expectation value $O(\epsilon^6)$ in the ground state.  We expect that if this sum-of-squares is equal to $H$ plus scalar, up to error $O(\epsilon^3)$, then we can adjust our trial Hamiltonian so that the sum-of-squares is exactly equal to $H$ plus scalar, at only an $O(\epsilon^6)$ error in energy.  A route to finding a sum-of-squares decomposition where terms have expectation value $O(\epsilon^6)$ would be to use the ``dressed operators" of \cref{dress} to construct the sum-of-squares.  This requires using dressed operators which are degree $5$, and hence a sum-of-squares of degree $10$.
The most optimistic situation at higher order would be that we could reproduce perturbation theory of order $2k+1$ using a degree-$(4k+2)$ sum-of-squares.

It would also be useful for quantum chemistry applications to extend the method to handle cases in which density-density interactions are strong.
In this case, our unperturbed Hamiltonian should have both $n_i$ and $n_i n_j$ terms of order unity, and our low-order trial Hamiltonian should allow arbitrary $n_i$ and $n_i n_j$ terms of order unity.  We will address this in a future work.

\appendix
\section{Implementation}
\label{implement}
This section contains a Python implementation of the algorithm of \cref{algo}.
It uses numpy's einsum to compute tensor contractions, and we use the ``optimal" argument to compute a best contraction method.
The implementation contains two functions, {\bf heffguide} and {\bf hselfconguide}.  The term ``guide" refers to the use of a guide Hamiltonian.

The function {\bf heffguide} computes $H_{eff}$, given a quadratic term $H_0$, quartic term $G$, number of orbitals $no$, and guide Hamiltonian $h0guide$.
Everywhere in this code, quadratic Hamiltonians are stored as $n_o$-by-$n_o$ matrices and quartic Hamiltonians are stored as four-index tensors, corresponding to the conventions at the start of \cref{numcon}.

The function {\bf heffguide} first computes $\tau$, storing its coefficients in tensor $T$ and matrix $T1$.  Then it computes $H_{\psi\tau}$, stored in $hpsitau$.
To compute $\tau$, it computes an array of Booleans, $excite$, with four entries, where each entry determines whether one of the four fermion operators in the quartic perturbation defined by $G$ will create an excitation (meaning, create a particle in a state with positive eigenvalue for $H_0$ or destroy one in a state with negative eigenvalue) or destroy an excitation (the opposite case to creating an excitation).  The variables plusenergy and minusenergy contain, respectively, the sum of absolute values of single particle eigenvalues for the four operators which create excitations and destroy excitations; whether or not the first operator $\psi^\dagger_{i,\sigma}$ in the term creates or destroys an excitation determines which of these we use to define $\tau_i$.
 
The function {\bf heffguide} returns six variables.  The first three, $scalarshift, h0shiftoldbasis, Gshiftoldbasis$, are, respectively, a scalar, quadratic, and quartic difference between $H_{eff}$ and the trial Hamiltonian.  The term ``oldbasis" is present in the variable names as this is after going back from the basis which diagonalizes the guide Hamiltonian to the ``old basis" in which the terms were given to this function.  The next three, $occupiedoldbasis, emptyoldbasis, h0guideoldbasis$ contain, respectively, projectors onto the occupied and empty orbitals of the new value of $h0guide$, as well as the new $h0guide$ itself, all in the old basis.

The function {\bf hselfconguide} contains a self-consistent loop which repeatedly calls {\bf heffguide} until the error is sufficiently small.
The error is computed as the sum of absolute values of coefficients of the linear and quartic terms in the difference between $H'$ and the target.

\begin{python}
import numpy as np
import functools

einsum=functools.partial(np.einsum,optimize='optimal')

def heffguide(h0,G,no,h0guide):
	no2=no*no
#initialize shift terms.
	h0shift=np.zeros((no,no))
	Gshift=np.zeros((no,no,no,no))
	scalarshift=0

#get eigenvalues and eigenvectors of h0
	h0guidevals,h0guidevecs = np.linalg.eigh(h0guide)	

#get G in new basis
	Gnew=einsum('abcd,ae,bf,cg,dh->efgh',G,h0guidevecs,h0guidevecs,h0guidevecs,h0guidevecs)

#effective Hamiltonian in terms of psi and tau.  Rows correspond to operators that create one particle, columns to those that destroy one
#first no entries are psi (or psi dagger), next no entries are tau (or tau dagger)
	hpsitau=np.zeros((2*no,2*no))
	for i in range(no):
		hpsitau[i,i+no]=h0guidevals[i]
		hpsitau[i+no,i]=h0guidevals[i]
	hpsitau[:no,:no]=einsum('ab,ae,bf->ef',h0,h0guidevecs,h0guidevecs)
	
#overlap matrix
	Spsitau=np.zeros((2*no,2*no))

#T will contain cubic terms in tau
	T=np.zeros((no,no,no,no))
	excite=[False,False,False,False]
	termvals=[0,0,0,0]
	for i in range(no):
		for j in range(no):
			for k in range(no):
				for l in range(no):
					plusenergy=0
					minusenergy=0
					excite[0]=(h0guidevals[i]>0)
					excite[1]=(h0guidevals[j]<0)
					excite[2]=(h0guidevals[k]>0)
					excite[3]=(h0guidevals[l]<0)
					termvals[0]=np.abs(h0guidevals[i])
					termvals[1]=np.abs(h0guidevals[j])
					termvals[2]=np.abs(h0guidevals[k])
					termvals[3]=np.abs(h0guidevals[l])
					nexcite=0
					for count in range(0,4):
						if excite[count]:
							nexcite+=1
							plusenergy+=termvals[count]
						else:
							minusenergy+=termvals[count]
					if nexcite==2:
						factor=0.5
					else:
						factor=1
					val=0
					if excite[0] and nexcite>=2:
						T[i,j,k,l]+=factor*Gnew[i,j,k,l]/plusenergy
					if not excite[0] and nexcite<=2:
						T[i,j,k,l]-=factor*Gnew[i,j,k,l]/minusenergy
					if excite[2] and nexcite>=2:
						T[k,l,i,j]+=factor*Gnew[i,j,k,l]/plusenergy
						val=factor*Gnew[i,j,k,l]/plusenergy
					if not excite[2] and nexcite<=2:
						T[k,l,i,j]-=factor*Gnew[i,j,k,l]/minusenergy
						val=-factor*Gnew[i,j,k,l]/minusenergy
					if j==k:
						hpsitau[i,l]+=h0guidevals[k]*val
						hpsitau[l,i]+=h0guidevals[k]*val
					if i==l:
						hpsitau[k,j]-=h0guidevals[k]*val
						hpsitau[j,k]-=h0guidevals[k]*val

#T1 will include all terms with single fermi operator in tau.  T and T1 both annihilate one particle
	T1=np.zeros((no,no))
	for i in range(0,no):
		for j in range(0,no):
			for k in range(0,no):
				for l in range(0,no):
						if k==l and h0guidevals[l]<0:
							T1[i,j]-=2*T[i,j,k,l]
						if j==k and h0guidevals[k]>0:
							T1[i,l]-=T[i,j,k,l]

#add terms from  T1 to hpsitau
	for i in range(0,no):
		for j in range(0,no):
			hpsitau[i,j]-=h0guidevals[i]*T1[i,j]
			hpsitau[j,i]-=h0guidevals[i]*T1[i,j]

	h0temp=np.copy(hpsitau[:no,:no])
	for i in range(0,no):           
		for j in range(0,no):   
			h0temp[i,j]+=hpsitau[i,i+no]*T1[i,j]
			h0temp[j,i]+=hpsitau[i,i+no]*T1[i,j]

	occupied=np.zeros((no,no))
	empty=np.zeros((no,no))
	for i in range(0,no):
		if h0guidevals[i]<0:
			occupied[i,i]=1
		if h0guidevals[i]>0:
			empty[i,i]=1
#build overlap matrix
	for i in range(0,no):
		Spsitau[i,i]=1
#compute tau tau dagger anticommutator
	tautauquart=einsum('abcd,jblm->jadclm',T,T)-0.5*einsum('abcd,jklb->jadclk',T,T)+0.5*einsum('abcd,jkcm->jadmbk',T,T)-0.5*einsum('abck,jklm->jalmbc',T,T)-0.5*einsum('abcm,jklm->jabklc',T,T)
	tautauquad=einsum('abcd,jkkb->jadc',T,T)-einsum('abcd,jbcm->jadm',T,T)+einsum('abbk,jklm->jalm',T,T)-0.5*einsum('abbm,jklm->jalk',T,T)+einsum('abcm,jblm->jalc',T,T)+einsum('ab,jblm->jalm',T1,T)-0.5*einsum('ab,jklb->jalk',T1,T)+einsum('abcd,jb->jadc',T,T1)-0.5*einsum('abcd,jd->jabc',T,T1)
	tautauscalar=einsum('abbm,jkkm->ja',T,T)+einsum('ab,jb->ja',T1,T1)+einsum('ab,jkkb->ja',T1,T)+einsum('abbd,jd->ja',T,T1)

	Stemp=tautauscalar+2*einsum('jkbc,bc->jk',tautauquad,occupied)+4*einsum('ab,cd,jkabcd->jk',occupied,occupied,tautauquart)+2*einsum('ac,bd,jkabdc->jk',occupied,empty,tautauquart)
	Spsitau[no:,no:]=np.copy(Stemp)
#Diagonalize Stemp. Change Spsitau so that block is diagonal.  Similarly, rotate tautau and rotate T and T1 (used below in getting psidagtau, etc...).  Then rotate hpsitau.  Then proceed as below.
	Svals,Svecs=np.linalg.eigh(Stemp)
	Stemp=einsum('ab,ac,bd->cd',Stemp,Svecs,Svecs)
	Spsitau[no:,no:]=np.copy(Stemp)
	tautauquart=einsum('abcdef,ag,bh->ghcdef',tautauquart,Svecs,Svecs)
	tautauquad=einsum('abcd,ag,bh->ghcd',tautauquad,Svecs,Svecs)
	tautauscalar=einsum('ab,ag,bh->gh',tautauscalar,Svecs,Svecs)
	T=einsum('abcd,ae->ebcd',T,Svecs)
	T1=einsum('ab,ae->eb',T1,Svecs)
	hpsitau[no:,:no]=einsum('ab,ac->cb',hpsitau[no:,:no],Svecs)
	hpsitau[:no,no:]=einsum('ab,bc->ac',hpsitau[:no,no:],Svecs)

#compute anticommutator of psi dagger tau
	psidagtauquad=einsum('abcd->bacd',T)-0.5*einsum('abcd->dacb',T)
	psidagtauscalar=einsum('abbd->da',T)+np.transpose(T1)
#compute anticommutator of tau dagger psi
	taudagpsiquad=einsum('abcd->abdc',T)-0.5*einsum('abcd->adbc',T)
	taudagpsiscalar=einsum('abbd->ad',T)+T1

#diagonalize hpsitau using overlap matrix
	for i in range(0,no):
		for j in range(0,no):
			hpsitau[j,i+no]*=np.sqrt(Spsitau[i+no,i+no])
			hpsitau[i+no,j]*=np.sqrt(Spsitau[i+no,i+no])
	psitauvals,psitauvecs=np.linalg.eigh(hpsitau)

#compute energy of hpsitau
	energypsitauham=0
	for i in range(0,2*no):
		if psitauvals[i]<0:
			energypsitauham+=2*psitauvals[i]
	print("energy of psitau hamiltonian is ",energypsitauham)

#undo scaling
	for i in range(0,2*no):
		for j in range(no,2*no):
			psitauvecs[j,i]/=np.sqrt(Spsitau[j,j])

#compute shifts
	for i in range(0,2*no):
		if psitauvals[i]<0:
			scalarshift-=2*psitauvals[i]*einsum('a,a',psitauvecs[:no,i],psitauvecs[:no,i])
			scalarshift-=2*psitauvals[i]*einsum('ab,a,b',tautauscalar,psitauvecs[no:,i],psitauvecs[no:,i])
			scalarshift-=2*psitauvals[i]*(einsum('ab,a,b',psidagtauscalar,psitauvecs[:no,i],psitauvecs[no:,i])+einsum('ab,a,b',taudagpsiscalar,psitauvecs[no:,i],psitauvecs[:no,i]))
			h0shift-=2*psitauvals[i]*einsum('abcd,a,b',tautauquad,psitauvecs[no:,i],psitauvecs[no:,i])
			h0shift-=2*psitauvals[i]*(einsum('abcd,a,b',psidagtauquad,psitauvecs[:no,i],psitauvecs[no:,i])+einsum('abcd,a,b',taudagpsiquad,psitauvecs[no:,i],psitauvecs[:no,i]))

			Gshift-=2*psitauvals[i]*einsum('abcdef,a,b->cdef',tautauquart,psitauvecs[no:,i],psitauvecs[no:,i])

#return Gshift and h0shift to original basis

	Gshiftoldbasis=einsum('abcd,ea,fb,gc,hd->efgh',Gshift,h0guidevecs,h0guidevecs,h0guidevecs,h0guidevecs)
	h0shiftoldbasis=einsum('ab,ea,fb->ef',h0shift,h0guidevecs,h0guidevecs)
	occupiedoldbasis=einsum('ab,ea,fb->ef',occupied,h0guidevecs,h0guidevecs)
	emptyoldbasis=einsum('ab,ea,fb->ef',empty,h0guidevecs,h0guidevecs)

#make explicitly hermitian.  The hamiltonian actually is hermitian but it has been written in a possibly nonhermitian way, with the quartic and quadratic terms each nonhermitian but the nonhermiticity canceling
	h0shiftoldbasis=0.5*h0shiftoldbasis+0.5*np.transpose(h0shiftoldbasis)
	Gshiftoldbasis=0.5*Gshiftoldbasis+0.5*einsum('abcd->dcba',Gshiftoldbasis)
	h0guideoldbasis=einsum('ab,ea,fb->ef',hpsitau[:no,:no],h0guidevecs,h0guidevecs)

	return scalarshift,h0shiftoldbasis,Gshiftoldbasis,occupiedoldbasis,emptyoldbasis,h0guideoldbasis

def selfconguide(a,G,no):
	f=0.5
	hscalar=0
	hscalartry=0
	Gtry=G
	h0guide=a
	h0try=a
	for counter in range(0,40):
		guidevals,guidevecs=np.linalg.eigh(h0guide)
		scalarshift,h0shift,Gshift,occupied,empty,guidenew=heffguide(h0try,Gtry,no,h0guide)
		h0guide=f*guidenew+(1-f)*h0guide
		energy0=scalarshift
		energy2=2*einsum('bc,bc',h0shift,occupied)
		energy4=4*einsum('ab,cd,abcd',occupied,occupied,Gshift)+2*einsum('ac,bd,abdc',occupied,empty,Gshift)
		print(energy0,energy2,energy4,"estimated total energy shift is",energy0+energy2+energy4-hscalartry)
		hscalartrynew=hscalar+energy2+energy4
		h0trynew=a-h0shift
		Gtrynew=G-Gshift
		err=np.abs(hscalartry-hscalartrynew)
		for i in range(0,no):
			for j in range(0,no):
				err+=2*np.abs(h0trynew[i,j]-h0try[i,j])
		for i in range(0,no):
			for j in range(0,no):
				for k in range(0,no):
					for l in range(0,no):
						err+=4*np.abs(Gtrynew[i,j,k,l]-Gtry[i,j,k,l])
		print("conservative error estimate is",err)
		
		hscalartry=hscalartrynew
		h0try=f*h0trynew+(1-f)*h0try
		Gtry=f*Gtrynew+(1-f)*Gtry
		if err<1e-5:
			break
	return(energy0+energy2+energy4-hscalartry,err)
\end{python}
\section{Dressed Fermionic Operators}
\label{dress}
To construct higher-order perturbation theory, it will be useful to construct operators $\tp_i$ which (approximately) annihilate the (perturbed) ground state $\gs$.
For $\epsilon=0$, we will have $\tp_i=\psi_i$, and they will be constructed perturbatively in $\epsilon$.  

We will say that an operator $\tp_i$ is a ``dressed" fermionic operator if $\tp_i$ if
$$\tp_i=\psi_i+O(\epsilon),$$
where the $O(\epsilon)$ term is some power series in $\epsilon$ and the operators $\psi_i,\psi^\dagger_i$.

In some particular cases, the operators $\tp_i$ will anti-commute with each other.  In this case, we say that $\tp_i$ obey the anti-commutation relations.  In some cases, we will also have the property that $\{\tp_i,\tp_j^\dagger\}=\delta_{i,j}$; in this case, we will say that $\tp_i,\tp_j^\dagger$ obey the anti-commutation relations.

\subsection{Perturbative Calculation of Dressed Operators To Second Order}
To construct these operators, one approach would be to construct, perturbatively in $\epsilon$, a unitary $U$ which maps the unperturbed ground state wavefunction $\Psi_0(0)$ to the perturbed wavefunction $\gs$, and then define $\tp_i=U \psi_i U^\dagger$.  Then, the series for $U$ gives a series for $\tp_i$ and the operators $\tp_i$ obey the anti-commutation relations, though any truncation of $\tp_i$ to finite order in $\eps$ need not obey these relations.

We can also construct the operators more directly.  In this subsection, we again include quadratic terms in the perturbation to see how the dressed operator changes under these terms.

Recall that
\begin{align}
V=&\Bigl(\sum_{i<j<k<l} V_{ijkl} \psi_i \psi_j \psi_kj \psi_l+\hc.\Bigr) + \Bigl(\sum_{i} \sum_{j<k<l} V_{\overline i j k l} \psi^\dagger_i \psi_j \psi_k \psi_l+\hc\Bigr)
\\ \nonumber &+ \sum_{i>j} \sum_{k<l} V_{\overline i \overline j k l} \psi^\dagger_i \psi^\dagger_j \psi_k \psi_l.
\end{align}

We now construct $\tp_i$ perturbatively in $\epsilon$.  We write $$\tp_i=\psi_i+\epsilon O^{(1)}_i + \epsilon^2 O^{(2)}_i + \ldots.$$
We will calculate $O^{(1)},O^{(2)}$ explicitly.  An important fact of this explicit calculation is that we need to only include terms up to degree $5$ in $O^{(2)}$ as terms of degree $7$ will cancel.  This is a result of a general principle\cite{hastings2022perturbation}, that we may choose terms proportional to $\epsilon^j$ in $\tp_j$ to be of degree at most $2j+1$.

We choose $O^{(1)}$ so that $O^{(1)}\Psi_0(0)$ cancels the term proportional to $\epsilon$ in $\psi_i \gs$, picking
$$O^{(1)}=-\sum_{j<k<l}
\frac{V_{\overline i \overline j \overline k \overline l}}{e_i+e_j+e_k+e_l} \psi^\dagger_j \psi^\dagger_k \psi^\dagger_l.$$

We now choose $O^{(2)}$ so that $O^{(2)} \Psi_0$ cancels the terms proportional to $\epsilon^2$
 in $(\psi_i + \epsilon O^{(1)}_i) \gs$.
 There are two sources of such terms, either from $\epsilon O^{(1)}_i$ acting on terms of order $\epsilon$ in $\gs$ or from
 $\psi_i$ acting on terms of order $\epsilon^2$ in $\gs$.

 The operator $\epsilon O^{(1)}_i$ acting on terms of order $\epsilon$ in $\gs$ gives 
$$\epsilon^2 \sum_{j<k<l}
\frac{V_{\overline i \overline j \overline k \overline l}}{e_i+e_j+e_k+e_l} \psi^\dagger_j \psi^\dagger_k \psi^\dagger_l
\sum_{m<n<o<p}
\frac{V_{\overline m \overline n \overline o \overline p}}{e_m+e_n+e_o+e_p} \psi^\dagger_m \psi^\dagger_n \psi^\dagger_o \psi^\dagger_p \Psi_0(0).$$
 This vanishes unless $j,k,l,m,n,o,p$ are all distinct.
 Suppose that $i$ however is not distinct from all $m,n,o,p$.  Suppose then, without loss of generality, that $i=m$ (other cases can be reduced to this by re-ordering of indices and possibly a sign change).
 One may verify then that
 $$\sum_{j<k<l}
\frac{V_{\overline i \overline j \overline k \overline l}}{e_i+e_j+e_k+e_l} \psi^\dagger_j \psi^\dagger_k \psi^\dagger_l
\sum_{n<o<p}
\frac{V_{\overline i \overline n \overline o \overline p}}{e_m+e_n+e_o+e_p} \psi^\dagger_m \psi^\dagger_n \psi^\dagger_o \psi^\dagger_p \Psi_0(0)+ jkl \leftrightarrow nop=0,$$
where the second term denotes the effect of interchanging $j,k,l$ with $n,o,p$.
So, we may assume that
$i,j,k,l,m,n,o,p$ are all distinct.
In this case, however, one may verify that for any given $i,j,k,l,m,n,o,p$ this term cancels exactly with $\psi_i$ acting on the sum of the two terms in $\gs$ which are of order $\epsilon^2$ in $\gs$ and which are proportional to
$V_{\overline i \overline j \overline k \overline l}V_{\overline m \overline n \overline o \overline p}$, namely
$$\epsilon^2 \psi_i 
\frac{V_{\overline i \overline j \overline k \overline l}}{e_i+e_j+e_k+e_l+e_m+e_n+e_o+e_p}
\psi^\dagger_i \psi^\dagger_j \psi^\dagger_k \psi^\dagger_l 
\frac{V_{\overline m \overline n \overline o \overline p}}{e_m+e_n+e_o+e_p}
\psi^\dagger_m \psi^\dagger_n \psi^\dagger_o \psi^\dagger_p \Psi_0(0)+ ijkl \leftrightarrow mnop.$$
So, indeed, all terms in $(\psi_i + \epsilon O^{(1)}_i) \gs$ with $7$ excitations above $\Psi_0$ (i.e. which are given by $7$ creation operators acting on $\Psi_0(0)$) cancel.

If we allow $V$ to include terms of degree $6$ or higher, similar cancellations happen so
that all terms in $(\psi_i + \epsilon O^{(1)}_i) \gs$ in which the perturbation enters in the form
$V_{\overline{\ldots}} V_{\overline{\ldots}}$ will vanish.  Here $V_{\overline{\ldots}}$ means set of indices all with overlines, i.e., some term in $V$ which is a sum of products of creation operators only.

Now,
let us consider
 $\psi_i$ acting on terms of order $\epsilon^2$ in $\gs$, considering all terms which we have not already used to cancel $\epsilon O^{(1)}_i$ acting on terms of order $\epsilon$ in $\gs$.  
 As noted, the second term of order $\epsilon^2$ in \cref{standard2pert} vanishes, and $\psi_i$ annihilates the last term.  So, we consider only the first term of order $\epsilon^2$ in \cref{standard2pert}, considering the case where the final eigenstate $\Psi_j$ has fewer than $8$ excitations above $\Psi_0$.  Acting with $\psi_i$ on these terms we get the state
 $\epsilon^2 O^{(2)} \Psi_0$
 for
 \begin{align}
 \label{O2}
 O^{(2)}=&-\Bigl(
 \sum_{j<k} \sum_m
 \sum_{n<o<p}
 \frac{V_{\overline i \overline j \overline k m}}{e_i+e_j+e_k+e_n+e_o+e_p}
 \frac{V_{\overline m \overline n \overline o \overline p}}{e_m+e_n+e_o+e_p}
\psi^\dagger_j \psi^\dagger_k  \psi^\dagger_n \psi^\dagger_o \psi^\dagger_p
  \\ \nonumber
 &+
 \sum_{j} \sum_{m<n}
 \sum_{o<p}
 \frac{V_{\overline i \overline j m n}}{e_i+e_j+e_o+e_p}
 \frac{V_{\overline m \overline n \overline o \overline p}}{e_m+e_n+e_o+e_p}
 \psi^\dagger_j \psi^\dagger_o \psi^\dagger_p
 \\ \nonumber
 &+  \sum_{m<n<o}
 \sum_{p}
 \frac{V_{\overline i m n o}}{e_i+e_p}
 \frac{V_{\overline m \overline n \overline o \overline p}}{e_m+e_n+e_o+e_p}
 \psi^\dagger_p
 \\ \nonumber
&-  \sum_{j<k<l} \sum_m
 \sum_{o<p}
 \frac{V_{\overline j \overline k \overline l m}}{e_i+e_j+e_k+e_l+e_o+e_p}
 \frac{V_{\overline i \overline m \overline o \overline p}}{e_i+e_m+e_o+e_p}
\psi^\dagger_j \psi^\dagger_k  \psi^\dagger_l \psi^\dagger_o \psi^\dagger_p
\\ \nonumber
 &+
 \sum_{j<k} \sum_{m<n}
 \sum_{p}
 \frac{V_{\overline j \overline k m n}}{e_i+e_j+e_k+e_p}
 \frac{V_{\overline m \overline n \overline i \overline p}}{e_m+e_n+e_i+e_p}
 \psi^\dagger_j \psi^\dagger_k \psi^\dagger_p
 \\ \nonumber
 &- \sum_{j} \sum_{m<n<o}
 \frac{V_{\overline j m n o}}{e_i+e_j}
 \frac{V_{\overline m \overline n \overline o \overline i}}{e_m+e_n+e_o+e_i}
 \psi^\dagger_j
 \Bigr).
 \end{align}
The first term of order $\epsilon^2$ in \cref{standard2pert} has $V$ appear twice in it; 
 the first three terms on the right-hand side of \cref{O2} correspond to cases where $i$ is in the first appearance of $V$ and the last three terms correspond to cases where it is in the second appearance.

One may then verify that if we define
\begin{align}
\tp_i =& \psi_i + \epsilon O^{(1)} + \epsilon^2 O^{(2)}
+\cO(\epsilon^3),
\end{align}
then this annihilates $\gs$.

There is of course some large arbitrariness in the choice of dressed operator if we impose only the requirement that it annihilate the perturbed ground state to given order in perturbation theory.
Given any choice of dressed operators $\tp_i$, then another choice for which the $\tp_i$ obey the anti-commutation relations is to take operators
$\exp(O ) \tp_i \exp(-O) $ for some operator $O$ which is a polynomial in the $\tp_i$.

\subsection{General Construction Size Consistency}
\label{sizecon}

The cancellation of terms with $7$ creation operators above is related to a property of size consistency.  Indeed, the self-consistent method as outlined here is size-consistent, in that given two Hamiltonians, $H_1$ and $H_2$, acting on different systems, then when the method is applied to the Hamiltonian $H_1+H_2$ on the combined system, the bound it gives is the sum of the bounds for $H_1$ and $H_2$ separately.

On the other hand, the 2RDM method is not always size consistent, if one imposes an additional global constraint on the number operators\cite{nakata2009size}.  This of course is not a drawback of the method: rather it is an advantage that one can impose a strict constraint on the number, and it would be useful to be able to do the same here.

\bibliography{qsos-ref}
\end{document}